# Mycorrhizal association of common European tree species shapes biomass and metabolic activity of bacterial and fungal communities in soil

Petr Heděnec[a1*], Lars Ola Nilsson[a1], Haifeng Zheng[a], Per Gundersen[a], Inger Kappel Schmidt[a]

Johannes Rousk[b] and Lars Vesterdal[a]

[a]Department of Geosciences and Natural Resource Management, University of Copenhagen, Rolighedsvej 23, DK-1958 Frederiksberg C, Denmark

[b]Department of Biology, Microbial Ecology - MEMEG, Lund University, Ecology Building, SE-223 62 Lund, Sweden

[1]Authors equally contributed to manuscript

* Corresponding author: **Petr Heděnec, PhD**

Email: peh@ign.ku.dk

Tel: +420731019906

Department of Geosciences and Natural Resource Management, University of Copenhagen, Rolighedsvej 23, DK-1958 Frederiksberg C, Denmark

**Highlights:**

- The soil microbial communities differed between six European tree species
- The EcM tree species had higher fungal growth and biomass
- The AM species showed higher bacterial growth
- Tree species effects were driven by litter quality and changed soil properties

**ABSTRACT**

Recent studies have revealed effects of various tree species on soil physical and chemical properties. However, effects of various tree species on composition and activity of soil microbiota and the relevant controls remain poorly understood. We evaluated the influence of tree species associated with two different mycorrhizal types, ectomycorrhiza (EcM) and arbuscular mycorrhiza (AM), on growth, biomass and metabolic activity of soil fungal and bacterial communities using common garden tree species experiments throughout Denmark. The soil microbial communities differed between six European tree species as well as between EcM (beech, lime, oak and spruce) and AM (ash and maple) tree species. The EcM tree species had higher fungal biomass, fungal growth and bacterial biomass, while AM species showed higher bacterial growth. The results indicated that microbial community composition and functioning differed between groups of tree species with distinct litter qualities that generate soil C/N ratio and soil pH differences. The mycorrhizal association only partly explained litter quality and soil microbial species differences since lime was more similar to AM tree species. In addition, our results indicated that tree species-mediated soil pH and C/N ratio were the most important variables shaping microbial communities with a positive effect on bacterial and a negative effect on fungal growth rates. The results suggest that tree species-mediated microbial community composition and activity may be important drivers of the different vertical soil C distribution previously observed in AM and EcM tree species.

# 1. Introduction

Recent studies indicated significant effects of aboveground vegetation on soil physical and chemical soil properties (Alvarez et al., 2009; De Schrijver et al., 2012; Vesterdal et al., 2008). For example, the C stocks and vertical distribution of C and N content in the soil profile differ between temperate and boreal tree species (Boča et al., 2014; Vesterdal et al., 2013). However, our understanding of tree species effects on the composition and activity of soil microbiota and the relevant controls still entails several challenges, particularly related to interactions between mycorrhizal fungi and microorganisms (Prescott and Grayston, 2013).

The biomass, growth and the composition of the microbial community is often predominantly shaped by soil pH in forest soils (Cruz-Paredes et al., 2017; Hobbie, 2015; Nilsson et al., 2007; Rousk et al., 2010a). Soil microbiota are also affected by land use history, e.g., forests of long continuity are likely to sustain higher biomass and diversity of soil microbiota than forests recently established on former agricultural soils (Moore-Kucera and Dick, 2008). However, other studies (Birkhofer et al., 2012; Evans et al., 2014; Grayston and Prescott, 2005) revealed that effects of other factors such as litter chemistry, soil moisture conditions and nutrient availability can overrule the effect of soil pH on microbial communities when tree species are grown on similar soils.

The chemical composition of leaf litter substantially affects composition and activity of the microbial community. Litter of "low quality" with high C/N ratio, low concentrations of other nutrients and high amounts of recalcitrant compounds (allelopathic compounds, tannins, lignin etc.) supports a fungal-dominated decomposition channel with slow nutrient turnover (Cornwell et al., 2008; Wardle et al., 2004). In contrast, "high quality" litter with low C/N ratio, high concentrations of other nutrients and low concentrations of recalcitrant compounds supports a bacterial-dominated decomposition channel with fast nutrient turnover (Wardle et al., 2004). In addition, soils dominated by trees producing low quality litter harbor gram-positive bacteria capable of degrading recalcitrant compounds while soils dominated by trees producing high quality litter are dominated by gram-negative bacteria decomposing labile compounds (Fanin et al., 2019; Kramer and Gleixner, 2008).

Two dominant classes of mycorrhizal fungi, the arbuscular mycorrhizal fungi (AM) and ectomycorrhizal fungi (EcM), form symbiosis with the roots of most trees on Earth, enhancing access to soil nutrients and water (van der Heijden et al., 2015). The AM fungi primarily scavenge for inorganic forms of N and P released by saprotrophic microbes, while EcM fungi rely on organic N and P sources which require extracellular enzymes that allow degrading complex organic N and P-bearing compounds in soil, such as proteins chitin and inositol phosphates (Phillips et al., 2013; Smith and Read, 2008; Wooliver et al., 2019).

The EcM fungi are the most important symbionts of temperate forest trees (Harley and Harley, 1987). They contribute largely to soil microbial biomass and also account for up to one third of tree fine root biomass (Högberg et al., 2007; Wallander et al., 2001). In contrast, AM fungi are ubiquitous among non-woody plant species, but also found on some temperate deciduous trees such as ash (*Fraxinus* sp.) and maple (*Acer* sp.). The dominant type of mycorrhizal symbiosis in different ecosystems is expected to shape the microbial communities to a high degree, most likely as a result of different nutrient economies established in soils under the two types of tree species (Craig et al., 2018; Phillips et al., 2013).

The more nutrient-rich litter (N, P and base cations) in AM associated tree species support an inorganic nutrient economy with higher soil pH stimulating a bacteria-dominated microbial community (Lin et al., 2017; Phillips et al., 2013). On the other hand, EcM associated tree species support an organic nutrient economy because of their ability to mine nutrients in organic matter by production of specific exoenzymes (Lin et al., 2017), thereby being less dependent on saprotrophic microbes for nutrient release (Phillips et al., 2013). The resultant large pool of organic N and acid soil in EcM tree species select for a fungal-dominated microbial community (Phillips et al., 2013). Altogether, these differences between AM and EcM associated tree species suggest that microbial communities and their activities would be more dominated by bacteria under AM species, whereas EcM species would have relatively higher fungal biomass and activity because of specific environmental conditions established as a result of contrasting leaf litter qualities and contrasting interactions between the two mycorrhizal types and saprotrophic microbial communities.

In the present study, we evaluated the influence of six common European tree species with different mycorrhizal associations on soil microbial communities in a 30-year-old common garden experiment that controlled for possible confounding site-related effects (Vesterdal et al., 2013). This replicated common garden experiment has previously revealed tree species effects on soil C and N stocks (Vesterdal et al., 2008), soil respiration and C and N turnover (Vesterdal et al., 2012), $^{15}$N abundance and N cycling traits (Callesen et al., 2013), and on N and water balances (Christiansen et al., 2010). We aimed to extend the understanding of plant-soil interactions by combining existing data on tree species-specific litter quality and soil physico-chemistry with (new) information on microbial community composition and function.

The specific objective was to quantify the effect of EcM and AM tree species with variable leaf litter traits on fungal and bacterial biomass and growth along with a range of microbial community composition indices. We hypothesized that i) the mycorrhizal associations and their inherent nutrient economies would shape the biomass and growth of soil bacteria and fungi, with higher bacterial biomass and activity under AM tree species and higher fungal biomass and activity under EcM tree species, and ii) that the tree species-mediated physico-chemical soil properties would explain differences in bacterial and fungal communities.

## 2. Material and methods

### 2.1. Site description and sampling design

The common garden experiments included 10 tree species randomly planted in 1973 in single-species plots (0.25 ha) in a block design replicated at six locations in Denmark (Table 1). The six common European tree species studied were beech (*Fagus sylvatica* L.), pedunculate oak (*Quercus robur* L.), lime (*Tilia cordata* L.), sycamore maple (*Acer pseudoplatanus* L.), ash (*Fraxinus excelsior* L.), and Norway spruce (*Picea abies* (L.) Karst.). Ash was missing in one of the six sites (Vallø) due to failed establishment. Ash and maple have AM association whereas beech, oak, lime and Norway spruce all have EcM association (Harley and Harley, 1987). Four of the experimental sites were planted on old forest land (Odsherred, Vallø, Viemose, Wedellsborg) and two on former cropland (Kragelund, Mattrup) (Table 1). The sites are distributed throughout Denmark, but mainly situated on relatively nutrient-rich soils developed from till deposits and classified as Luvisols (Mattrup, Odsherred, Vallø, Viemose), Phaeozem (Wedellsborg) and Alisol (Kragelund) (Callesen et al., 2003). The soil texture of the parent material differed among sites, e.g. with clay content increasing from 8% (Kragelund) to 30% (Wedellsborg) in the C horizon (Vesterdal et al., 2008). The sites were mainly located in, or close to, rural and intensively managed agricultural areas with intermediate N deposition of 8-16 kg/ha/yr (J.L. Bak (pers. comm.) based on regional modeling of N deposition (Ellermann et al., 2018) adjusted for local agricultural emissions according to Bak et al. (2018). Climatic conditions regarding mean annual precipitation (580–890 mm), mean annual temperature (7.5-8.4°C) and the length of the growing season are relatively similar (Vesterdal et al., 2008). Litter turnover rates were high and there was only little forest floor accumulation (L layers) in plots with deciduous trees while forest floors under spruce had FH layers (Vesterdal et al., 2012). Further site information can be found in Vesterdal et al. (2008). Three soil samples were collected from each plot of the six tree species in each of the six sites in the autumn 2005 from 0-5 cm of the mineral soil using a soil corer (4 cm diameter). The three subsamples from each plot were pooled to one composite sample for each tree species at each site, so there were in total 35 samples (6 tree species x 6 sites, but ash missing at the site Vallø. The soil samples were sieved through a 2 mm mesh and stored at -20°C. After one month, the soil samples were thawed and pre-incubated for one week at 5 °C before analyses. All analyses started immediately after pre-incubation.

**Table 1**

Site and mineral soil (0-5 cm) characteristics (mean±SEM).

| Site | | Location | Former land-use | Established (year) | Soil type (WRB) | Soil pH ($H_2O$) | Organic matter (%) | Soil moisture (%) | Clay (%) | Silt (%) | Sand (%) |
|---|---|---|---|---|---|---|---|---|---|---|---|
| 1 | Kragelund | 56°10'N, 9°25'E | Cropland | 1961 | Arenic Alisol | 4.5±0.1bc | 5.3±0.8c | 18.7±1.4b | 8 | 7 | 85 |
| 2 | Mattrup | 55°570N, 9°38'E | Cropland | 1973-74 | Lamellic Luvisol | 5.0±0.3a | 6.0±0.3bc | 24.0±0.7ab | 17 | 9 | 74 |
| 3 | Odsherred | 55°50'N, 11°42'E | Old forest | 1973-74 | Stagnic Luvisol | 4.4±0.2c | 9.6±0.7 a | 26.2±1.7a | 11 | 11 | 78 |
| 4 | Vallø | 55°25'N, 12°03'E | Old forest | 1973-74 | Stagnic Luvisol | 4.4±0.3c | 12.1±2.4a | 25.7±1.1a | 17 | 13 | 70 |
| 5 | Viemose | 55°01'N, 12°09'E | Old forest | 1973-74 | Stagnic Luvisol | 4.5±0.1c | 8.3±0.5ab | 27.3±2.4a | 25 | 20 | 55 |
| 6 | Wedellsborg | 55°24'N, 9°52'E | Old forest | 1973-74 | Luvic Phaeozem | 4.9±0.3ab | 8.7±0.4ab | 25.4±2.0a | 30 | 30 | 40 |

Clay, <2 μm; silt, 2–20 μm; sand, 20–2000 μm)) in the 50–100 cm layer (Vesterdal et al., 2008). Data with the same or no letters do not differ significantly by Tukeys HSD pairwise comparisons. A priori planned comparisons were done between tree species with ectomycorrhiza (EcM) and arbuscular mycorrhiza (AM).

*2.2. Soil physico-chemical soil properties*

The soil moisture was measured gravimetrically after drying at 105 °C. The pH was measured at the soil:water ratio of 1:5 and analysed with a Radiometer combination-electrode GK2401 (Radiometer, Copenhagen, Denmark). Total C and N were analyzed as total C and N by dry combustion (Dumas method) in a Leco CSN 2000 Analyzer (Matejovic, 1993). There was no inorganic C ($CaCO_3$) within 50 cm depth in soils at the six sites (pH<5.5) and all measured C were consequently considered to be organic. Soil organic matter was measured by loss on ignition at 400°C.

*2.3. Analysis of microbial biomass and activity*

Microbial activity was estimated as basal respiration at 20°C, measured on a gas chromatograph (GC) after 24 h incubation of soils in 20 ml sealed glass vials, and microbial biomass was estimated from substrate induced respiration (SIR) two hours after glucose addition (Anderson and Domsch, 1978). The soil bacterial growth was estimated as the rate of leucine (Leu) incorporation into bacteria extracted from soil after homogenization and centrifugation (Baath, 1994; Bååth, 2001; Rousk and Bååth, 2011). The amounts of Leu incorporated into extracted bacteria $g^{-1}$ soil $h^{-1}$ were used as a measure of bacterial growth. The amount of acetate (Ac) incorporated into fungal ergosterol $g^{-1}$ soil $h^{-1}$ was used as a measure of fungal growth (Bååth, 2001). The fungal/bacterial growth ratio was calculated as the rate of fungal growth vs. bacterial growth. Fungal biomass was estimated assuming 5 mg ergosterol $g^{-1}$ fungal biomass (Joergensen, 2000). The ergosterol peak was collected and the amount of incorporated activity was determined.

The soil microbial community composition, including bacterial and fungal biomass, was estimated from the phospholipid fatty acid (PLFA) patterns of soil samples (Frostegård et al., 1993). Nomenclature of PLFAs is according to Frostegård et al. (1993). The PLFA 18:2ω6,9 was used as an indicator of fungal biomass and PLFAs for estimation of bacterial biomass included i14:0, 14:0, i15:0, a15:0, 15:0, i16:0, 16:1ω7t, br17:0, i17:0, a17:0, cy17:0, br18:0, 10Me17:0, 18:1ω7, 10Me18:0, 19:1a and cy19:0. The fungal to bacterial ratio was achieved from the 18:2ω6,9 to bacterial PLFA ratio (Frostegård et al., 2011). The ratio between Gram-positive ($G^+$) and Gram-negative ($G^-$) bacteria (the $G^+/G^-$ ratio) was based on the sum of the PLFAs i14:0, i15:0, a15:0, i16:0, br17:0, i17:0, a17:0, 10Me17:0 and br18:0 as indicators of Gram-positive ($G^+$) bacteria, and the sum of the PLFAs cy17:0, 18:1ω7 and cy19:0 for Gram-negative ($G^-$) bacteria. The sum of actinomycetes included 10Me16:0, 10Me17:0 and 10Me18:0.

## *2.4. Statistical analyses*

The influence of tree species on microbial community characteristics was analyzed using one-way analysis of variance (GLM) of tree species effects. Site was included in the analysis as a block factor because there was just one plot of individual tree species within a site. The experimental design did therefore not allow to address possible tree species x site interaction. The significance levels were: $P<0.1$ (trend) and $P<0.05$ (significance). Post-hoc pairwise comparisons were done mainly by Tukey's HSD, but LSD results are reported in case of statistical trends ($0.05<P<0.1$). A priori planned comparisons were used to evaluate the effect of the mycorrhizal association of the tree species with AM (ash and maple) vs EcM tree species (beech, lime, oak and spruce), alternatively the deciduous EcM species (beech, lime and oak) hereafter denoted as EcM species. The PLFA data (relative mol % of the 30 PLFAs) were analysed by principal component analyses (PCA) with potentially explanatory environmental data as supplementary variables, not included in the PCA, but as overlays to the factor scores. The first and second principal components (PC1 and PC2) were used in further analyses using regression models. Pearson's correlation coefficient was used to investigate correlation between soil physico-chemical properties and fungal and bacterial biomass and growth. All statistical tests were carried out using STATISTICA version 8.0 (StatSoft®).

## 3. Results

### *3.1. Effects of EcM and AM tree species on physico-chemical soil properties*

Soils below AM tree species had higher pH than soils planted by tree species associated with EcM type (Table 2). Soils planted by Norway spruce were lowest in pH (mean 4.0), beech and oak were intermediate (4.3-4.4), while soils planted by lime, maple and ash were highest in pH (4.9-5.2). In contrast, soils planted by EcM trees had higher organic matter concentrations and C/N ratios than soils beneath AM trees (Table 2). The tree species ranking in C/N ratio was opposite to that for pH (spruce highest, beech and oak intermediate and ash, maple and lime lowest). Organic matter concentrations were highest beneath Norway spruce but only differed significantly from maple. Norway spruce soils were drier than soils beneath all four deciduous species.

**Table 2**

Tree species-specific soil properties (mean±SEM) in the mineral soil (0-5 cm).

| Tree species | Mycorrhizal type | Soil pH (H$_2$O) | Organic matter LOI (%) | C/N ratio | Soil moisture (weight %) |
|---|---|---|---|---|---|
| Norway spruce | EcM | 4.0±0.1$^d$ | 10.9±2.1$^a$ | 17.7±1.3$^a$ | 18.7±1.2$^b$ |
| Beech | EcM | 4.3±0.1$^{cd}$ | 7.3±0.9$^{ab}$ | 15.1±0.7$^b$ | 26.1±2.1$^a$ |
| Oak | EcM | 4.4±0.1$^{bcd}$ | 8.3±0.9$^{ab}$ | 15.0±0.7$^b$ | 26.4±0.8$^a$ |
| Lime | EcM | 4.9±0.2$^{abc}$ | 8.3±0.9$^{ab}$ | 14.5±0.8$^{bc}$ | 25.4±1.1$^a$ |
| Maple | AM | 5.0±0.2$^{ab}$ | 7.0±0.8$^b$ | 13.7±0.5$^{cd}$ | 25.2±2.1$^a$ |
| Ash | AM | 5.2±0.3$^a$ | 7.5±1.6$^{ab}$ | 13.0±0.6$^d$ | 25.4±2.6$^a$ |
| GLM | | *** | (*) | *** | ** |
| Planned comp. | | AM > EcM | AM < EcM | AM < EcM | AM > EcM |

Two-way GLM analyses were performed with site and species as variables, without an interaction term. Significance levels: 0.1 ((*)), 0.05 (*), 0.01 (**) and 0.001 (***). Data with different lettering differ significantly in Tukeys HSD (if not otherwise stated) pairwise comparisons. A priori planned comparisons were done between tree species with ectomycorrhiza (EcM) and arbuscular mycorrhiza (AM).

## 3.2. Effects of EcM and AM tree species on biomass and growth of fungal and bacterial community

Soil microbial biomass and growth as well as biomass growth of fungi and bacteria were strongly affected by tree species and their mycorrhizal association. The EcM species had higher fungal biomass, fungal activity, fungal/bacterial growth ratio, $G^+/G^-$ ratio and higher proportions of actinomycetes, while AM species had higher microbial biomass and bacterial growth (Fig. 1, Table 3). Among the EcM species, lime was an outlier with higher microbial biomass and bacterial activity, but lower fungal activity and lime was more similar to the AM species in these microbial traits.

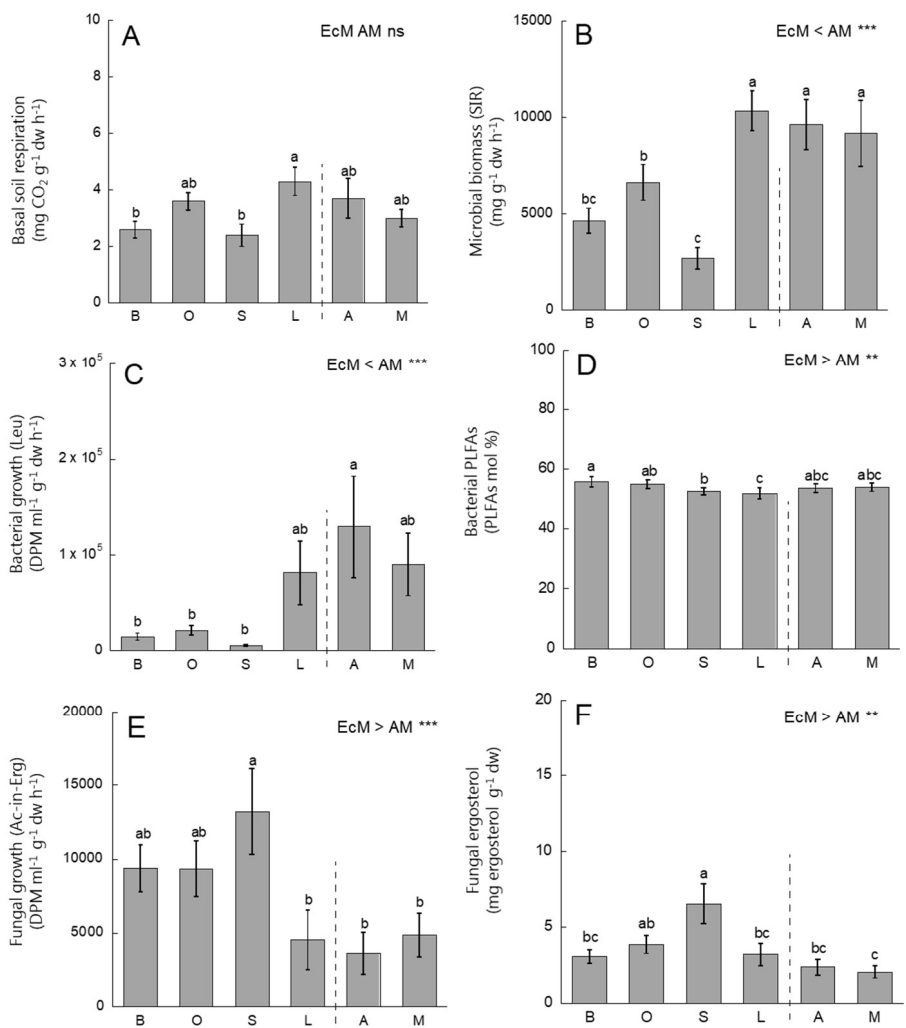

**Fig. 1.** Soil microbial variables in forest soils with different tree species (B = beech, O = oak, S = spruce, A = ash, L = lime, M = maple). Mean±SE values across six sites of microbial activity (basal respiration) **(A)**,

microbial biomass (SIR) (**B**), bacterial activity (leucine incorporation) (**C**), bacterial biomass (mol % bacterial PLFAs) (**D**), fungal activity (ac-in-erg) (**E**) and fungal biomass (ergosterol) (**F**). Differences between trees with EcM (B L O S) and AM (A M) symbiosis were tested by Tukey test. Asterisks *, ** and *** indicates significance levels of $P$< 0.05, 0.01 and 0.001 respectively.

Among individual tree species, lime, maple and ash were generally higher than beech, oak and spruce in microbial biomass and bacterial growth but lower in fungal growth, although differences were not always significant between all species (Fig. 1B, C, E and Table 3). Soils beneath spruce had highest fungal growth but lowest bacterial growth. In contrast, ash soils had highest bacterial growth but lowest fungal growth. Consequently, there was a marked increase in fungal/bacterial growth ratio in the order ash < maple < oak, beech < spruce. The fungal/bacterial ratio found in lime was similar to ratios in maple and oak (Table 3). Microbial activity was significantly higher in lime than beech and spruce. Beech was highest and lime lowest in bacterial biomass while spruce showed highest and maple and ash lowest fungal biomass (Fig 1D-F). The fungal biomass based on PLFA 18:2 ω 6,9 was also highest in spruce but lowest in oak and maple (Table 3). The two fungal abundance estimates were only weakly positively correlated ($R^2$=0.13; P=0.03). Tree species effects on all individual PLFAs are given in Table S2 (Supporting Information). The fungal to bacterial PLFA ratio was higher in spruce and lime than in beech and oak, and actinomycetes were relatively most abundant under spruce and least under lime, maple and ash (Table 3). The $G^+/G^-$ ratios were higher in spruce than in all five deciduous species (Table 3).

**Table 3**

Microbial community composition based on PLFA analysis (mean±SEM) for tree species in the common garden experiments.

| Tree species | Fungal PLFAs [1] (mol %) | Fungal / Bacterial ratio [2] | Fungal/ Bacterial growth rate | Actinomycetes PLFAs[3] (mol %) | $G^+ / G^-$ ratio [4] ratio |
|---|---|---|---|---|---|
| Norway spruce | 2.5±0.3$^a$ | 0.049±0.006$^a$ | 2.73±0.63$^a$ | 12.5±1.0$^a$ | 2.01±0.11$^a$ |
| Beech | 1.7±0.2$^{ab}$ | 0.030±0.004$^b$ | 0.77±0.16$^b$ | 8.5±0.5$^b$ | 1.24±0.11$^b$ |
| Oak | 1.6±0.2$^b$ | 0.030±0.004$^b$ | 0.73±0.32$^b$ | 7.1±0.8$^{bc}$ | 1.16±0.13$^b$ |
| Lime | 2.5±0.3$^{ab}$ | 0.048±0.008$^a$ | 0.28±0.23$^{cb}$ | 6.4±0.7$^c$ | 1.02±0.09$^b$ |
| Maple | 1.7±0.2$^{ab}$ | 0.032±0.004$^{ab}$ | 0.19±0.10$^c$ | 6.7±0.4$^{bc}$ | 1.02±0.05$^b$ |
| Ash | 2.1±0.5$^{ab}$ | 0.040±0.011$^{ab}$ | 0.07±0.04$^d$ | 6.1±0.4$^c$ | 1.18±0.15$^b$ |
| GLM | ** | ** | *** | *** | *** |
| Planned comp. | EcM >AM | EcM >AM | EcM >AM | EcM >AM | EcM>AM |

For explanation of abbreviations and statistical analysis see Table 1. When ratios are used, and when Levene's test of homogeneity of variances or Brown & Forsythe's test of the homogeneity of variances <0.05, tests were done using log-transformed data. Data with the same or no letters do not differ significantly by Tukeys HSD pairwise comparisons. A priori planned comparisons were done between tree species with ectomycorrhiza (EcM) and arbuscular mycorrhiza (AM).

[1] Mol % PLFA 18:2ω6,9 divided with all included PLFAs

[2] Sum of bacterial PLFA (i14:0 + 14:0 + i15:0 + a15:0 + 15:0 + i16:0 + 16:1ω7t + br17:0 + i17:0 + a17:0 + cy17:0 + br18:0 + 10Me17:0 + 18:1ω7 + 10Me18:0 + 19:1a + cy19:0) mol % divided with all included PLFAs.

[3] Sum of actinomycete PLFA (10Me16:0 + 10Me17:0 + 10Me18:0) mol % divided with all included PLFAs.

[4] Ratio (i14:0 + i15:0 + a15:0 + i16:0 + br17:0 + i17:0 + a17:0 + 10Me17:0 + br18:0) / (cy17:0 +, 18:1ω7 + cy19:0)

*3.3. Effects of former land use on biomass and growth of fungal and bacterial communities*

Two of the six common garden sites had an agricultural land-use history, and we found differences in biomass and growth of fungal and bacterial communities between the former land uses (Table S1). Soils in sites with long forest continuity (former forest land, Table 1) showed higher bacterial and fungal biomass and growth than soil former agricultural lands (Fig. S1A-F).

*3.4. Effects of tree species-mediated soil physico-chemical properties on the soil microbial community composition and activity*

The first and second principal components in the PC analyses accounted for 39.4 and 17.2% of the variation, respectively. PC1 was mainly associated with tree species-related variation in litter quality and topsoil chemistry (Table 2), i.e. negatively related to foliar litter C/N ratio ($R^2=0.24$; $P<0.01$) and soil C/N ratio ($R^2=0.31$; $P<0.001$) and positively related to soil pH ($R^2=0.54$; $P<0.001$) (Fig. 2D). PC2 mainly represented site-related variation ($P<0.001$), but also represented a minor component of variation among the five deciduous tree species ($P<0.05$). PC2 was positively related to e.g. soil pH ($R^2=0.18$; $P<0.05$) and negatively related to soil organic matter ($R^2=0.12$; $P<0.01$), C and N concentrations ($R^2=0.32-0.39$; $P<0.001$; Fig. 2D), and soil moisture ($R^2=0.12$; $P<0.01$).

EcM and AM species were not completely separated along PC1, as the ordination plot revealed convergence of lime with ash, and maple also showed convergence with oak and beech along PC2 (Fig. 2A). Tree species separated clearly along PC1 (spruce << beech < oak < lime = maple = ash), but deciduous tree species were also separated along PC2 (beech = oak < lime with maple and ash intermediate) (Fig. 2A). The PCA analyses revealed clear divergence between former cropland and former forested sites along PC2 (Fig 2B) with higher pH and lower organic matter, C and concentrations at former cropland sites.

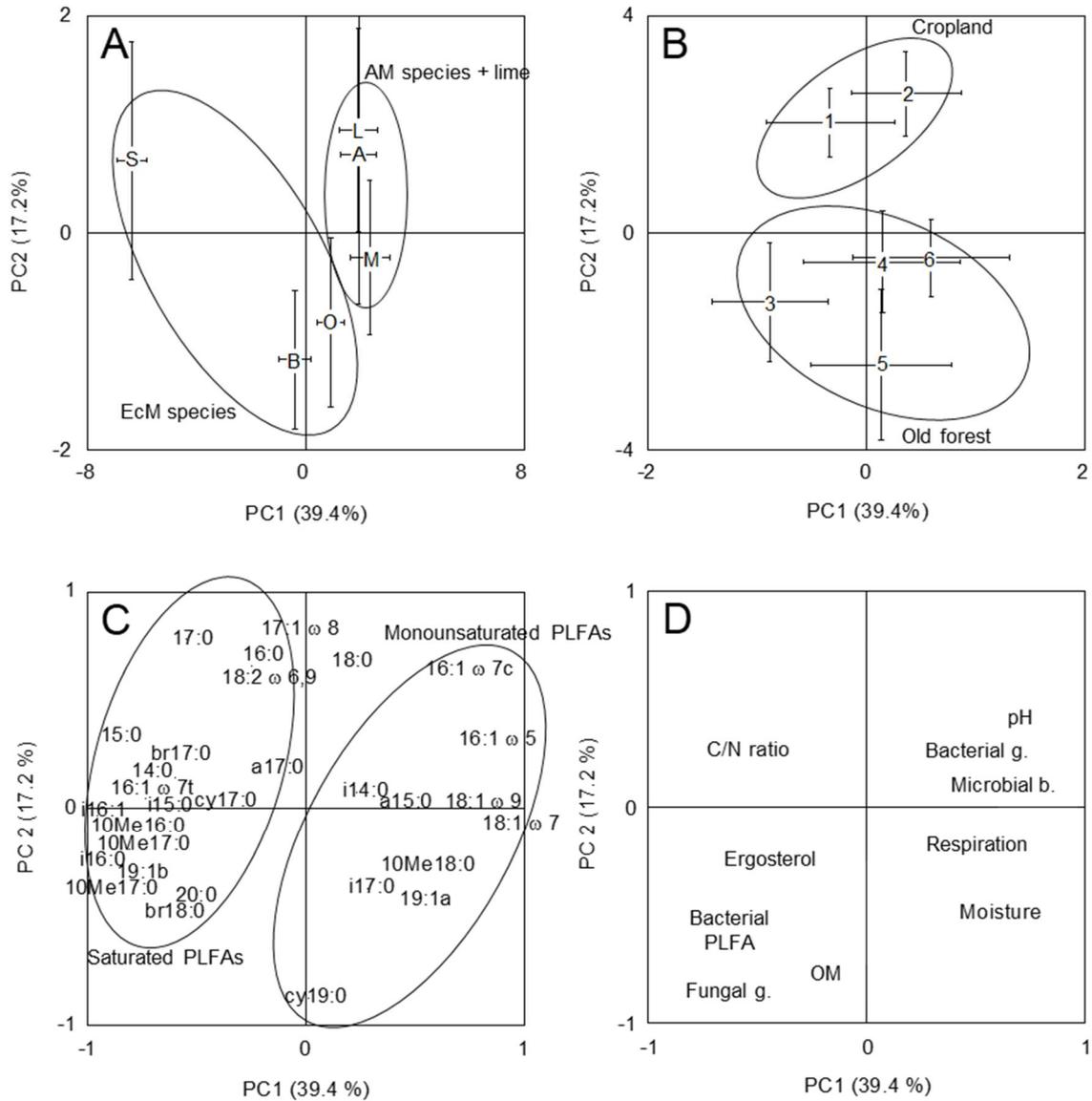

**Fig. 2.** PCA plots for soil microbial and physico-chemical properties (0-5 cm) in the common garden experiment with six tree species at six sites. PCA scores from the different (A) tree species and (B) sites (mean values ± s.e.). (C) PC loadings of the individual PLFAs and (D) environmental variables and microbial biomass (b.) and growth (g.) indices.

For the microbial variables (Fig. 2D), PC1 (associated with tree species-mediated litter and soil N and soil pH) was positively related to microbial biomass and growth and to bacterial growth but negatively related to fungal growth and fungal biomass (ergosterol) (Table 4). Fungal/bacterial growth ratios were closely negatively associated with PC1. Fungal growth and bacterial PLFA were also negatively related to PC2 (site and former land-use related pH and C, N and OM concentrations), indicating higher values in old forest sites.

The ordination plot indicated that PLFAs separated in two groups (Fig. 2C). One group represented monounsaturated PLFAs such as PLFAs 18:1ω7, 16:1 ω 5, 18:1 ω 9 and 16:1ω7c whereas the other group of saturated PLFAs included PLFAs i16:1, i16:0, 10Me16:0 and 10Me17:0 (Fig. 2C). The saturated PLFAs were associated with EcM species (particularly spruce) that were conducive to drier soils with high C/N ratio and low pH, whereas monounsaturated PLFAs were associated with AM species that had more moist soils with lower C/N ratio and higher pH (Fig. 2C). Actinomycetes and the $G^+/G^-$ ratio were strongly negatively related to PC1 (Table 4), i.e. they were high at low bacterial growth and increased with fungal growth. The $G^+/G^-$ ratio was also negatively related to PC2 indicating higher ratios in old forest sites.

**Table 4**

Linear regressions between principal components representing soil physico-chemical properties between various tree species (PC1) and sites across Denmark (PC2) and microbial characteristics.

| Microbial characteristics | PC1 (Tree species effect) | | | PC2 (Site effect) | | |
|---|---|---|---|---|---|---|
| | Pearson r | $R^2$ | p value | Pearson r | $R^2$ | p value |
| Microbial biomass (SIR) | 0.77 | 0.60 | <0.001 | 0.04 | 0.01 | 0.802 |
| Basal soil respiration | 0.37 | 0.13 | 0.03 | -0.18 | 0.07 | 0.129 |
| Fungal biomass (ergosterol) | -0.62 | 0.38 | <0.001 | -0.15 | 0.02 | 0.38 |
| Fungal growth DW | -0.65 | 0.43 | <0.001 | -0.46 | 0.21 | 0.01 |
| Bacteria (PLFA)[1] | 0.06 | 0.004 | 0.71 | -0.85 | 0.72 | <0.001 |
| Fungi (PLFA)[2] | 0.61 | 0.37 | <0.001 | 0.70 | 0.49 | <0.001 |
| Bacterial growth DW | 0.55 | 0.30 | <0.001 | 0.14 | 0.02 | 0.41 |
| Fungal/bacterial ratio | 0.39 | 0.16 | 0.017 | 0.80 | 0.65 | <0.001 |
| Fungal/bacterial growth | -0.80 | 0.65 | <0.001 | -0.19 | 0.03 | 0.26 |
| Actinomycetes (PLFA)[3] | -0.88 | 0.78 | <0.001 | -0.14 | 0.02 | 0.42 |
| $G^+/G^-$ [4] ratio | -0.84 | 0.70 | <0.001 | 0.37 | 0.14 | 0.03 |

[1] Sum of bacterial PLFA (i14:0 + 14:0 + i15:0 + a15:0 + 15:0 + i16:0 + 16:1ω7t + br17:0 + i17:0 + a17:0 + cy17:0 + br18:0 + 10Me17:0 + 18:1ω7 + 10Me18:0 + 19:1a + cy19:0) mol % divided with all included PLFAs.

[2] Mol % PLFA 18:2ω6,9 divided with all included PLFAs

[3] Sum of actinomycete PLFA (10Me16:0 + 10Me17:0 + 10Me18:0) mol % divided with all included PLFAs.

[4] Ratio (i14:0 + i15:0 + a15:0 + i16:0 + br17:0 + i17:0 + a17:0 + 10Me17:0 + br18:0) / (cy17:0 +, 18:1ω7 + cy19:0)

## 4. Discussion

### *4.1 Effect of tree species and type of mycorrhiza*

We found a consistent pattern in microbial communities and their activities in AM and EcM tree species. Microbial SIR-biomass, bacterial biomass (PLFA) and bacterial growth were lower, while fungal biomass (ergosterol) and growth as well as fungi/bacteria ratio and fungal/bacterial growth ratios were higher in soils beneath EcM tree species than in soils beneath AM tree species. Moreover, EcM species hosted bacterial communities that were more dominated by actinomycetes and $G^+$ bacteria. These results corroborated the hypothesis that soil microbes and their activities under EcM trees would differ from trees associated with AM types. This was not only driven by a conifer-broadleaf contrast but was also found within the deciduous tree species. The effects of tree species on soil microbes mirror the patterns of C and N cycling and storage in the same and other common garden experiments (De Schrijver et al., 2012; Schelfhout et al., 2017; Vesterdal et al., 2012, 2008). Specifically, ash, maple (AM) and lime (EcM) had high turnover rates of litter, N-rich, lignin-poor litter and high soil pH, while spruce (EcM) had low soil pH, N-poor and lignin-rich litter, low litter turnover rates and high accumulation of C and N in the forest floor rather than in the mineral soil. Beech and oak (EcM) had characteristics that often were intermediate between ash, maple and lime vs. spruce.

Consistent with our results, a global meta-analysis of EcM vs. AM tree species (Lin et al., 2017) found similar contrasts in leaf litter C/N and lignin concentrations and in topsoil C/N ratio between mycorrhizal types but higher microbial activity in AM species was only translated into higher N mineralization and nitrification rates and not higher respiration and C mineralization rates.

We also expect effects of mycorrhizal type on microbial activity via the roots, i.e. related to the input of organic matter and its quality for microbes. However, there is not sufficient evidence for differences in root litter or exudate input between EcM and AM tree species (Lin et al., 2017), and information on root litter quality is scarce and cannot be inferred from quality of leaf litter (Hobbie et al., 2010). In boreal forests, a major pathway for C input to soils was reported to occur via ectomycorrhizal mycelia (Clemmensen et al., 2013), but the difference to AM tree species is not well known. A few studies reported higher root litter quality in AM trees (Makita and Fujii, 2015; Taylor et al., 2016), which suggest stimulation of bacterial processing also through the roots. On the other hand, Phillips and Fahey (2006) showed greater positive effect of EcM than AM tree species on microbial biomass and mineralization in rhizosphere compared to bulk soil. The relative importance of the roots warrants further investigation to explain effects of tree species on microbial communities.

The higher microbial activities in AM species are in line with observations of Craig et al. (2018) that growth and turnover of microbial biomass is greater under AM tree species. These results altogether provide new and strong evidence that AM species among the most common European tree species are conducive to a more active microbial community with a higher abundance of bacteria, and that this is linked to differences in C-nutrient couplings as suggested by Phillips et al. (2013). The common EcM species, on the other hand, hosted a higher share of actinomycetes and other bacterial groups which were reported to be more stress-tolerant toward moisture, C and N availability (Brockett et al., 2012; Fanin et al., 2019). The EcM species have lower quality litter in terms of C/N ratio, lignin, P and base cation concentrations (Langenbruch et al., 2012; Vesterdal et al., 2012) which leads to more stressful environments in terms of C and nutrient availabilities. Accordingly, high abundance of actinomycetes and high $G^+/G^-$ ratio were reported to indicate environmental stress in microbial communities experiencing drought (Grayston and Prescott, 2005). Fanin et al. (2019) further demonstrated that gram-negative bacteria are associated with simple C compounds (alkyls), whereas gram-positive bacteria are strongly associated with more complex C compounds such as carbonyls. These findings match well with the high $G^+/G^-$ ratio and high abundance of actinomycetes in spruce stands where drought is more frequent because of higher interception evaporation (Christiansen et al., 2010). As lignin concentrations and lignin/N ratio (as proxies for complex C compounds and N availability) were higher in litter of spruce, beech, oak and lime than that of ash and maple (Vesterdal et al., 2012) our study supports that the $G^+/G^-$ ratio can be employed as a useful indicator of the relative C availability for soil bacterial communities (Kramer and Gleixner, 2008).

Another important factor behind the effect of tree species on biomass and activity of soil microbiota could be presence and activity of earthworms. For example, a study by Goméz-Brandón et al. (2012) showed that the type of earthworm-induced modifications of the soil microbiota depend on the type and quality of substrate ingested. In addition, a recent study by Heděnec et al. (2020) revealed that presence of earthworms increased microbial activity, biomass and nutrient transformations in laboratory microcosms. Earthworms were previously studied in the same common garden experiment, and earthworm abundance was higher in AM tree species and lime, than in the remaining EcM tree species (Schelfhout et al., 2017). This coincidence between high earthworm and bacterial abundances in AM tree species and lime support that earthworms could be mediating the tree species effect on microbial communities.

Lime was the only exception to the clear separation in microbial indices between AM and EcM tree species. Lime trees (*Tilia cordata* Mill.) are considered to be associated with EcM, but can also be associated with AM fungi (Harley and Harley, 1987). We found lime to have some properties in common with EcM species such as high fungal biomass (from PLFAs) and high fungal/bacteria ratio. However, most attributes in lime stands were more typical for AM tree species such as ash and maple, i.e. high microbial activity and bacterial growth and low fungal growth, i.e., low fungal/bacterial ratio, and few actinomycetes. Lime has also previously been reported to have high soil pH and base saturation (Hagen-Thorn et al., 2004; Nordén, 1994a; Schelfhout et al., 2017) and high leaf litter nutrient concentrations (Desie et al., 2020; Langenbruch et al., 2012; Vesterdal et al., 2012), and the C,N and base cation cycling profile of lime is much closer to the AM tree species than to the deciduous EcM tree species oak and beech (Vesterdal et al., 2012, 2008). This suggests a strategy of lime with rapid inorganic nutrient cycling that contrasts with the typical strategy of EcM tree species (Phillips et al., 2013). The other broadleaf EcM species differed from lime by their lower soil pH, higher litter and soil C/N ratios, lower microbial biomass and bacterial activity, and slower litter turnover with accumulation of C in the organic horizon (Vesterdal et al., 2012, 2008). These findings are also consistent with results from mature natural forests, where Scheibe et al. (2015) found higher microbial biomass under the AM species ash and maple than EcM species beech and hornbeam, but that lime was intermediate. The closer functional similarity of lime with AM species may suggest that this tree species could differ in its mycorrhizal association (Pigott, 1991). However, the evidence of it being associated with EcM (Harley and Harley, 1987; Timonen and Kauppinen, 2008), and specifically so in a range of Danish conditions and forest types (Nielsen and Rasmussen, 1999), suggests that lime is a "functional outlier" among the included EcM tree species, and that microbial communities may be driven by phylogenetic tree species traits rather than the mycorrhizal association *per se* (Koele et al., 2012).

Fungal biomass and abundance as estimated by ergosterol and the PLFA 18:2ω6,9 was lower in AM species. The two estimates were only weakly correlated, which is not surprising since the PLFA 18:2ω6,9 targets only saprotrophic and EcM fungi (Olsson, 1999; Zhao et al., 2005). AM fungi do not contain the PLFA 18:2ω6,9 but contain 16:1ω5, which has previously been suggested as a signature lipid for AM fungi (Olsson, 1999). Consistent with this we found PLFA 16:1ω5 to be most abundant in AM tree species (Table S2). However, contributions from gram-negative bacteria can also contribute to higher abundance of PLFA 16:1ω5 at higher pH even in soils with plants that lack AM fungi (Rousk et al., 2010b, 2013). Fungal PLFA markers in mycorrhizal types were largely consistent with expectations, although high abundance of both AM fungi and gram-negative bacteria could be causing higher abundance of PLFA 16:1ω5 under ash and maple.

*4.2. Relationships with tree species –mediated soil properties*

Tree species define the quality of the organic matter for microbes, but an indirect pathway for effects on microbes occurs via their effects on the topsoil environment, i.e. the litter quality in terms of base cations and C/N ratio affects the chemistry of the mineral soil (De Schrijver et al., 2012; Desie et al., 2020; Finzi et al., 1998). We observed tree species-specific influences on topsoil chemistry (Table 2), which were reflected in the quality of the litter in terms of N, base cation concentrations and pH (Schelfhout et al., 2017; Vesterdal et al., 2008). Tree species strongly affected soil pH, C/N ratios and moisture, and these soil properties were shaping the microbial communities (Fig. 2). For example, close correlation between foliar litter N and base cation concentrations and soil C/N ratios and base cations, respectively, in the common garden (Schelfhout et al., 2017; Vesterdal et al., 2008) indicated an indirect role of litter N and base cations on microbial activities mediated through soil chemistry. Important roles of soil C/N ratios, especially for tree species related differences, and of soil pH, in addition to accumulation of organic matter, and land-use history were also shown.

The bacterial and fungal growth rates (Fig. 2C-D) were affected to a higher degree along the tree species-mediated soil gradient (PC1) than was basal respiration (i.e. C mineralization), which represent the combined functional contribution by both decomposer groups. This emphasizes the significance of separate bacterial and fungal growth measurements as early indicators of responses in soil microbial communities to changes in the environment.

The pronounced associations between tree species-related litter quality and soil chemistry and microbial indices supported our second hypothesis that litter quality-mediated physico-chemical soil properties would explain differences in bacterial and fungal communities. Our study of gradients in litter traits and soil nutrient status established by just two AM and four EcM tree species common to Europe provides evidence that the mycorrhizal-associated nutrient economy framework (Phillips et al., 2013), recently supported by global-scale meta-analysis of mycorrhizal types (Lin et al., 2017), holds truth under common garden conditions. We found that AM tree species formed the end of the soil gradient where a more inorganic nutrient economy was indicated by high pH and low C/N ratios, and where the microbial community was more bacteria-dominated and active. This is in line with mull humus forms and high rates of decomposition in AM tree species (Vesterdal et al., 2012, 2008) and corresponds with the strategy of AM fungi to scavenge for inorganic nutrients released by saprotrophic microbes (Phillips et al., 2013).

Ectomycorrhizal tree species, in particular spruce, had quite specific characteristics such as higher fungal activity, fungal biomass, $G^+/G^-$ ratio and more actinomycetes (Fig. 1; Table 2) which could be driven by lower soil pH. The shift in soil microbial community composition along the tree species-mediated soil gradient (PC1) was largely associated with PLFAs such as 16:1ω5, 18:1ω7 and 18:1ω9 (Fig. 2c), and these PLFAs are reported to correlate positively with pH (Nilsson et al., 2007; Thoms et al., 2010) and were thus more abundant in AM and lime stands. Bacterial PLFAs found at the other end of PC1, such as the PLFAs i16:0, 10Me16:0 and 10Me17:0 are associated with low soil pH (Nilsson et al., 2007).

Site differences in soil organic matter and former land-use also shaped the microbial communities, but this effect was weaker than that of tree species-mediated topsoil conditions (PC2, Fig. 2). The separation of sites according to former land use occurred even though the two former cropland sites differed in texture. The imprint of former cropland remained >30 years after afforestation in terms of lower fungal growth and bacterial PLFA, but higher fungal PLFA and fungi/bacteria ratio.

The pronounced tree species effects and the quite distinct separation of AM and EcM tree species on soil conditions and microbial indices are likely to translate into effects on soil C sequestration. The mull humus forms and fast forest floor turnover rates in AM tree species would indicate quick return of C to the atmosphere (Vesterdal et al., 2012), whereas EcM species (particularly conifers), with their pronounced accumulation of C in forest floors, would be more conducive to soil C sequestration (Berg, 2014; Lehmann and Kleber, 2015; Wiesmeier et al., 2019). More focus on mineral soil C and the role of microbiota have led to the hypothesis that tree species supporting more microbial and in particular bacterial processing of organic matter will result in more stable SOC in mineral soils (Cotrufo et al., 2013). The evidence is, however, still sparse. This and previous studies in the common garden experiment (Vesterdal et al., 2012, 2008) indeed suggest a positive association between higher microbial, and specifically bacterial, activity and higher mineral soil C stocks in the AM tree species ash and maple compared to EcM tree species. This link between microbial processing and soil C stocks have also been corroborated by studies along gradients of AM-EcM dominance in temperate North American forests (Craig et al., 2018). Further studies of the nexus between microbial communities and soil C stocks and forms should also involve interactions with soil fauna (Prescott and Grayston, 2013) because of higher earthworm abundance and activity observed in AM versus EcM tree species (Frouz, 2018; Schelfhout et al., 2017).

## 5. Conclusions

Soil microbial biomass and activity were primarily affected by tree species, and effects were related to litter quality as well as species-mediated soil characteristics. AM species (ash and maple) and EcM species (beech, lime, oak and spruce) differed in fungal and bacterial growth and biomass as well as actinomycetes and various bioindicators of microbial limitation. Forests with AM trees, with their high pH and nutrient concentrations in foliage and soil, high turnover rates and low fungal growth represent different traits than those with EcM trees. Lime trees were more related to AM trees in most respects, with high pH and turnover rates, high microbial biomass, bacterial biomass and growth, but had higher fungal PLFA in common with the other EcM species. Site conditions related to former land use (cropland versus old forest sites) still had some influence after 30 years in terms different SOM content that also influenced the soil microbial properties, however, to a lesser degree than tree species. Soil pH driven by differences in nutrient economy between AM and EcM tree species likely played an important direct or indirect role in shaping the microbial community activity and composition. However, soil pH correlated with several other variables, such as soil C/N ratio and litter C/N ratio and base cation content suggesting that soil microbiota biomass and growth are shaped directly by tree species-specific litter quality and indirectly through the soil chemistry. The results suggest that tree species-mediated microbial community composition and activity may be important drivers of different vertical soil C distribution previously observed in AM and EcM tree species.

**Acknowledgements**

This study was enabled through grants from the Swedish Research Council for Environment, Agricultural Sciences and Spatial Planning (Formas) and from Marie Curie under the 6th Framework Programme (EU) (grants to LON) and a grant by the Royal Physiographic Society of Lund (Kgl. Fysiografen) to JR. PH was supported by a Marie Curie Individual Fellowship (747824-AFOREST-H2020-MSCA-IF-2016/H2020-MSCA-IF-2016). We acknowledge China Scholarship Council for the PhD scholarship grant to HZ (201806910047).

**Mycorrhizal association of common European tree species shapes biomass and metabolic activity of bacterial and fungal communities in soil**

Petr Heděnec[a1*], Lars Ola Nilsson[a1], Haifeng Zheng[a], Per Gundersen[a], Inger Kappel Schmidt[a] Johannes Rousk[b] and Lars Vesterdal[a]

[a]Department of Geosciences and Natural Resource Management, University of Copenhagen, Rolighedsvej 23, DK-1958 Frederiksberg C, Denmark

[b]Department of Microbial Ecology, Lund University, Ecology Building, SE-223 62 Lund, Sweden

[1]Authors equally contributed to manuscript

* Corresponding author: Petr Heděnec

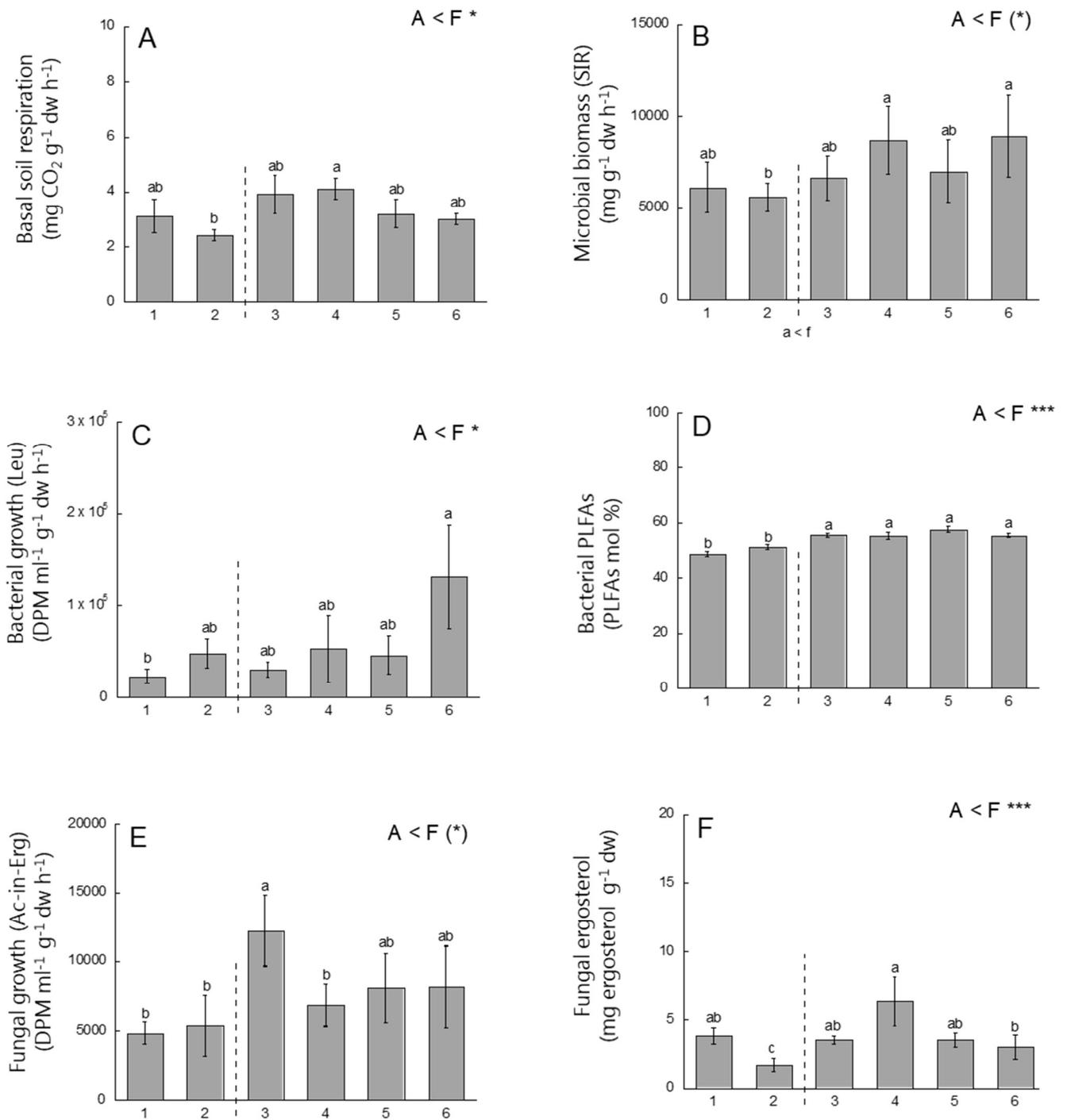

**Fig. S1.** Soil microbial variables in forest soils from different sites. Mean±SE values of microbial activity (basal respiration) (**A**), microbial biomass (SIR) (**B**), bacterial activity (leucine incorporation) (**C**), bacterial biomass (mol % bacterial PLFAs) (**D**), fungal activity (ac-in-erg) (**E**) and fungal biomass (ergosterol) (**F**). For explanation of site abbreviations see Table 1. Statistical tests to estimate the effect of site: two-way ANOVA (GLM) without an interaction term (so far on absolute, not transformed, data).

Table S1

Data from PLFA analysis and fungal/bacterial growth ratio (mean±SE) for various sites across Denmark.

|   | Site | Fungal biomass[1] (mol %) | Fungal / Bacterial[2] ratio | Fungal / bacterial growth ratio | Actinomycetes[3] (mol %) | G+ / G-[4] ratio |
|---|---|---|---|---|---|---|
| 1 | Kragelund | 2.8±0.3[a] | 0.058±0.007[a] | 0.5±0.3 | 7.8±0.9[ab] | 1.34±0.18[ab] |
| 2 | Mattrup | 2.1±0.4[ab] | 0.041±0.008[ab] | 0.7±0.4 | 7.3±0.9[ab] | 1.54±0.14[a] |
| 3 | Odsherred | 1.5±0.1[b] | 0.027±0.002[b] | 1.1±0.7 | 9.1±1.5[a] | 1.35±0.21[ab] |
| 4 | Vallø | 2.3±0.2[ab] | 0.043±0.005[ab] | 0.8±0.4 | 6.5±1.0[b] | 1.11±0.24[b] |
| 5 | Viemose | 1.7±0.2[b] | 0.030±0.004[b] | 1.0±0.7 | 8.6±1.4[ab] | 1.04±0.17[b] |
| 6 | Wedellsborg | 1.7±0.3[b] | 0.031±0.005[b] | 0.8±0.5 | 8.1±0.9[ab] | 1.24±0.09[ab] |
|   | GLM | ** | *** | ns | * | ** |
|   | Planned comp. | a > f |  |  |  | a > f |

Data with the same or no letters do not differ significantly by Tukeys HSD pairwise comparisons. A priori planned comparisons were done between tree species with ectomycorrhiza (EcM) and arbuscular mycorrhiza (AM).

Table S2

Individual PLFA abundance (mol %) in soil under different tree species.

|  |  | Ash | Beech | Lime | Maple | Oak | Spruce |
|---|---|---|---|---|---|---|---|
| i14:0 | ns | 1.8±1.1 | 1.7±0.4 | 1.6±1.1 | 1.6±0.4 | 1.3±0.3 | 1.4±0.4 |
| 14:0 | ** | 1.1±0.2 [b] | 1.2±0.3 [b] | 1.0±0.2 [b] | 1.1±0.2 [b] | 1.3±0.3 [b] | 1.7±0.3 [a] |
| i15:0 | * | 9.3±0.9 [ab] | 10.9±2.4 [ab] | 8.6±1.9 [b] | 8.8±1.3 [b] | 11.1±1.8 [ab] | 12.1±2.3 [a] |
| a15:0 | ns | 6.0±1.8 | 5.1±1.4 | 4.6±1.0 | 5.5±1.0 | 5.0±1.1 | 4.2±0.9 |
| 15:0 | *** | 0.6±0.1 [b] | 0.7±0.1 [b] | 0.6±0.1 [b] | 0.6±0.4 [b] | 0.7±0.1 [b] | 1.2±0.2 [a] |
| i16:1 | *** | 0.9±0.1 [b] | 1.1±0.2 [b] | 0.9±0.2 [b] | 0.9±0.4 [b] | 1.0±0.1 [b] | 1.9±0.3 [a] |
| i16:0 | *** | 3.4±0.5 [b] | 4.5±0.6 [b] | 3.4±0.7 [b] | 3.5±0.9 [b] | 3.9±0.3 [a] | 6.5±1.2 [a] |
| 16:1ω7c | ** | 8.1±1.2 [a] | 6.6±1.0 [ab] | 7.5±1.2 [a] | 7.8±1.3 [a] | 7.1±1.0 [ab] | 5.2±1.1 [b] |
| 16:1ω7t | *** | 1.0±0.1 [b] | 1.1±0.2 [b] | 0.9±0.1 [b] | 0.9±0.1 [b] | 1.0±0.2 [b] | 1.4±0.4 [a] |
| 16:1ω5 | *** | 4.3±0.1 [a] | 2.7±0.6 [b] | 3.9±1.0 [ab] | 4.5±1.0 [a] | 3.3±0.5 [ab] | 1.2±0.3 [c] |
| 16:0 | ns | 13.7±1.8 | 14.2±1.6 | 15.3±2.9 | 13.9±1.9 | 14.2±2.0 | 15.7±2.6 |
| br17:0 | * | 0.5±0.2 [ab] | 0.5±0.1 [ab] | 0.4±0.1 [ab] | 0.5±0.3 [ab] | 0.4±0.1 [b] | 0.7±0.2 [a] |
| 10Me16:0 | *** | 3.8±0.6 [b] | 5.8±1.4 [ab] | 4.0±1.5 [b] | 4.0±0.8 [b] | 4.7±1.9 [b] | 9.4±1.6 [a] |
| i17:0 | * | 1.8±0.2 [ab] | 1.9±0.2 [ab] | 1.7±0.1 [ab] | 1.9±0.2 [ab] | 1.9±0.2 [a] | 1.5±0.3 [b] |
| a17:0 | ns | 1.7±0.7 | 1.3±0.2 | 1.7±1.0 | 1.6±0.4 | 1.3±0.2 | 2.0±0.4 |
| 17:1ω8 | (*) | 0.5±0.2 | 0.4±0.1 | 0.5±0.1 | 0.4±0.1 | 0.4±0.1 | 0.5±0.1 |
| cy17:0 | ns | 2.8±0.4 | 2.9±0.5 | 2.7±0.3 | 2.5±1.1 | 2.7±0.6 | 3.1±0.4 |
| 17:0 | * | 0.5±0.2 [ab] | 0.5±0.1 [ab] | 0.5±0.1 [ab] | 0.4±0.1 [b] | 0.4±0.1 [b] | 0.7±0.1 [a] |
| br18:0 | * | 1.0±0.2 [ab] | 1.3±0.3 [ab] | 0.8±0.2 [b] | 0.9±0.3 [ab] | 1.0±0.3 [ab] | 1.5±0.7 [a] |
| 10Me17:0 | *** | 0.7±0.2 [b] | 1.1±0.2 [b] | 0.8±0.1 [b] | 0.8±0.3 [b] | 0.9±0.1 [b] | 1.8±0.5 [a] |
| 18:2ω6,9 | ns | 2.1±1.1 | 1.7±0.4 | 2.5±0.8 | 1.7±0.4 | 1.6±0.5 | 2.5±0.7 |
| 18:1ω9 | *** | 9.0±1.8 [a] | 7.4±1.2 [ab] | 9.6±1.5 [a] | 8.7±0.8 [a] | 8.5±1.9 [a] | 5.5±1.1 [b] |
| 18:1ω7 | *** | 13.4±1.9 [a] | 11.2±2.2 [a] | 13.2±2.1 [a] | 14.0±2.3 [a] | 12.2±2.1 [a] | 5.5±1.7 [b] |
| 18:0 | ns | 2.1±0.2 | 2.1±0.2 | 2.3±0.3 | 2.2±0.2 | 2.1±0.3 | 2.1±0.2 |
| 19:1a | ns | 0.9±0.4 | 0.9±0.1 | 0.8±0.2 | 0.9±0.2 | 0.9±0.2 | 0.6±0.2 |
| 10Me18:0 | ns | 1.6±0.3 | 1.6±0.4 | 1.6±0.4 | 1.9±0.5 | 1.5±0.2 | 1.3±0.4 |
| 19:1b | *** | 0.2±0.1 [b] | 0.4±0.1 [b] | 0.3±0.1 [b] | 0.3±0.2 [b] | 0.3±0.1 [b] | 0.8±0.3 [a] |
| cy19:0 | ns | 7.2±3.2 | 9.2±2.7 | 8.2±3.5 | 8.0±2.9 | 9.1±3.0 | 7.4±1.6 |
| 20:0 | ns | 0.4±0.1 | 0.5±0.1 | 0.4±0.1 | 0.4±0.1 | 0.4±0.1 | 0.5±0.3 |

Data with the same or no letters do not differ significantly by Tukeys HSD pairwise comparisons.